\documentclass[aps,pra,amsmath,amssymb,onecolumn,longbibliography]{revtex4-1}

\usepackage{graphicx}
\usepackage{subfig}
\usepackage{color}
\usepackage{stmaryrd}
\usepackage{epsfig}	
\usepackage{amsbsy}
\usepackage{dcolumn}
\usepackage{bm}
\allowdisplaybreaks

\begin{document}

\preprint{APS/123-QED}

\title[A closure theory for  the split energy-helicity cascades in homogeneous isotropic homochiral turbulence]{A closure theory for the split energy-helicity cascades
  in homogeneous isotropic homochiral turbulence}

\author{Antoine Briard $^1$}
\author{Luca Biferale$^2$}
\author{Thomas Gomez$^{3}$}
\email{thomas.gomez@univ-lille1.fr}

\affiliation{1- Sorbonne Universit\'es, UPMC Univ Paris 06, UMR 7190, Institut Jean Le Rond d'Alembert, F-75005, Paris, France}
\affiliation{2- Department of Physics and INFN, University of Rome 'Tor Vergata,' Via della Ricerca Scientifica 1, 00133 Rome, Italy}
\affiliation{3- Universit\'e Lille Nord de France, F-59000 Lille, France}

\date{\today}

\begin{abstract}
  We study the energy transfer properties of three dimensional homogeneous and isotropic turbulence where the non-linear transfer is altered in a way that helicity is made sign-definite, say positive. In this framework, known as homochiral turbulence, an adapted eddy-damped quasi-normal Markovian (EDQNM) closure is derived to analyze the dynamics at very large Reynolds numbers, of order $10^5$ based on the Taylor scale. In agreement with previous findings, an inverse cascade of energy with a kinetic energy spectrum like
  $\propto  k^{-5/3}$ is found for scales larger than the forcing one. Conjointly, a forward cascade of helicity  towards larger wavenumbers is obtained, where the kinetic energy spectrum scales like $\propto k^{-7/3}$.
By following the evolution of the closed spectral equations for a very long time and over a huge extensions of scales, we found the developing of a non monotonic shape for the front of the inverse energy  flux. The very long time evolution of the kinetic energy and integral scale in both the forced and unforced cases is analyzed also.
\end{abstract}

                              
\maketitle

\noindent
\section{Introduction}

Since the discovery that helicity is an inviscid invariant of the three-dimensional Navier-Stokes equations \cite{Moffatt1969}, the possibility of inverse cascades in homogeneous isotropic turbulence (HIT) without mirror symmetry has been greatly investigated: indeed, two-dimensional turbulence possesses as well two inviscid invariants, kinetic energy and enstrophy, and in this configuration, energy cascades towards large scales \citep{Boffetta2012}. 

In 3D, the first theoretical considerations date back to the study of Brissaud and coworkers \citep{Brissaud1973} where two different scenarios were proposed: (i) a joint direct  cascade of kinetic energy and helicity where $E(k) \sim \epsilon^{2/3} k^{-5/3}$, and $H(k) \sim \epsilon_H \epsilon^{-1/3}  k^{-5/3}$, where $E$ and $H$ are the kinetic and helical spectra, and $\epsilon$ and $\epsilon_H$ the kinetic energy and helicity dissipation rates; (ii) a split cascade with a direct transfer of helicity combined with an inverse cascade of kinetic energy. In fact, the second scenario was proven to be impossible by \cite{Andre1977} within the eddy-damped quasi-normal Markovian (EDQNM) approximation \citep{Orszag1970, Lesieur2008, Sagaut2008}. In subsequent papers like \cite{Borue1997,Chen2003} and more recently \cite{BG2017}, the joint direct cascade of helicity was observed  in HIT without mirror symmetry, with strictly zero  inverse energy transfer. This stems from the fact that helicity, the scalar product of velocity and vorticity, is not positive-definite, unlike kinetic energy and enstrophy.

Keeping this later feature in mind, Biferale and coworkers \citep{Biferale2012, Biferale2013} performed a "surgery" of the triadic Fourier interactions of turbulence, following the ideas developed by \cite{Waleffe1992}, in order to keep only the ones that maintain the helicity sign-definite, yielding to  the so-called homochiral turbulence. They consequently recovered the second scenario of \cite{Brissaud1973}, namely an inverse cascade of kinetic energy and a forward cascade of helicity, showing that all three dimensional turbulent flows indeed possess a sub-set of Fourier interactions potentially able to sustain an inverse energy cascade. Such a "surgery" of the Navier-Stokes equations was thoroughly investigated in multiple subsequent works \citep{Sahoo2015, Sahoo2017PRL, Sahoo2017PRF} so that the details are not recalled here. The main findings are that by forcing at small scales the decimated Navier-Stokes (dNS) equations where helicity is made sign-definite, say positive here, kinetic energy is transferred to smaller wavenumbers with $E(k)  \sim \epsilon^{2/3} k^{-5/3}$. 
By forcing at large scales the dNS equations, a direct helicity cascade with $E(k) \sim \epsilon_H^{2/3} k^{-7/3}$ was obtained. It was also notably shown that as soon as helicity is not made strictly sign-definite, by adding helical Fourier modes with the opposite helicity (negative here), the inverse energy cascade vanishes, and that the transition between the upward and forward cascades mechanisms looks singular \cite{Sahoo2015,Sahoo2017PRF}. On the other hand, by changing the relative weight of homochiral and heterochiral
triads, one is led to a transition from direct to inverse cascade for a finite value of the control parameter \cite{Sahoo2017PRL}, showing that the way Navier Stokes equations transfer energy across scales might be strongly different by changing the involved degrees of freedom.  \\

Strangely enough, inverse cascades with EDQNM were rarely investigated  in the past and mainly for  two configurations only: the inverse cascade of kinetic energy in 2D turbulence \citep{Pouquet1975}, and the inverse cascade of magnetic helicity in isotropic magnetohydrodynamics turbulence \citep{Pouquet1976}. Consequently, it appears interesting in terms of modelling to check that a sophisticated model such as EDQNM can handle complex and particular configurations when only specific triadic interactions are kept.
In the present work, we aim at further studying these features when helicity is made sign-definite by deriving a new EDQNM model. We show for the first time that
a split cascade scenario develops and that it can be studied for very  large Reynolds numbers and  for very long times.
In what follows, the adapted EDQNM closure for homochiral turbulence is derived, and then numerical results are presented for cases with and without forcing.

\section{EDQNM closure for homochiral turbulence}

The spectral counterpart of the Navier-Stokes equation for the  fluctuating velocity $\hat{u}_i$ reads in homogeneous isotropic turbulence 
\begin{align}
\left(\frac{\partial}{\partial t} + \nu k^2 \right) \hat{u}_j(\boldsymbol{k}) = - \mathrm{i} P_{jmn}(\boldsymbol{k}) \, \int_{\boldsymbol{k} = \boldsymbol{p} + \boldsymbol{q}} u_m(\boldsymbol{p}) u_n(\boldsymbol{q}) \mathrm{d}^3 \boldsymbol{p},
\label{eq_NS}
\end{align}
where $\nu$ is the kinematic viscosity, $\mathrm{i}^2=-1$, the operator $2 P_{jmn} = k_m P_{jn} + k_n P_{jm}$, with the projector $P_{jn} = \delta_{jn} - \alpha_j \alpha_n$ and $\alpha_j = k_j/k$, $k$ and $\boldsymbol{k}$ are respectively the wavenumber and wavevector, and $\hat{(\cdot)}$ denotes the Fourier transform. 
In isotropic turbulence  without reflexion symmetry, the spectral second-order velocity-velocity correlation
$\hat{R}_{ij}(\boldsymbol{k},t) = \langle\hat{u}_i^*(\boldsymbol{k},t) \hat{u}_j(\boldsymbol{k},t)\rangle$, where $(\cdot)^*$ denotes the complex conjugate and $\langle . \rangle$ an ensemble average, can be decomposed as 
\begin{equation}
\hat{R}_{ij}(\boldsymbol{k},t )  = \mathcal{E}(\boldsymbol{k},t) P_{ij}(\boldsymbol{k}) + \mathrm{i} \epsilon_{ijk} \alpha_k \frac{\mathcal{H}(\boldsymbol{k},t)}{k},
\label{EH_decomp}
\end{equation}
with $\epsilon_{ijk}$ the Levi-Civita permutation tensor, and where $\mathcal{E}$ and $\mathcal{H}$ are respectively the kinetic energy and kinetic helicity densities. 
Such a decomposition involving the kinetic energy density $\mathcal{E}$ and helical density $\mathcal{H}$ was previously used in \cite{Cambon1989,Borue1997,Chen2003}. More recently \cite{BG2017}, this decomposition was applied to investigate with EDQNM the large Reynolds numbers dynamics of the kinetic energy and helical spectra defined as
\begin{equation}
E(k,t) = \int_{S_k} \mathcal{E}(\boldsymbol{k},t) \mathrm{d}^2 \boldsymbol{k}, \qquad
H(k,t) = \int_{S_k} \mathcal{H}(\boldsymbol{k},t) \mathrm{d}^2 \boldsymbol{k},
\end{equation}
where $S_k$ is a sphere of radius $k$. The total helicity, which is the scalar product of velocity and vorticity, is then obtained by $\langle \boldsymbol{u}.\boldsymbol{\omega} \rangle/2 = \int_0^\infty H(k,t) \mathrm{d} k$.
In what follows, we analyze within an adapted EDQNM approximation at large Reynolds numbers, the  configuration proposed in \cite{Biferale2012,Biferale2013}, namely the homochiral \textit{decimated Navier-Stokes}.

To select only helical modes of identical sign the classical Lin equation for $E(k,t)$ \citep{Lin1949} and its non-linear transfer of HIT cannot be used anymore. One needs a new closure, adapted to homochiral turbulence where helicity is made sign-definite. First, the spectral fluctuating velocity is decomposed in positive and negative modes using the helical decomposition \citep{Cambon1989, Waleffe1992, Sagaut2008} so that 
\begin{equation}
\hat{\boldsymbol{u}}(\boldsymbol{k},t) = u_+(\boldsymbol{k},t) \boldsymbol{N}(\boldsymbol{k}) + u_- (\boldsymbol{k},t) \boldsymbol{N}^* (\boldsymbol{k}),
\end{equation}
where $\boldsymbol{N}$ are the helical modes, which verify $\boldsymbol{N}^*(\boldsymbol{k}) = \boldsymbol{N}(-\boldsymbol{k})$, $\boldsymbol{N}.\boldsymbol{N}=0$, $\boldsymbol{N} . \boldsymbol{N}^*=2$, and $ \mathrm{i} \boldsymbol{k} \times \boldsymbol{N} = k \boldsymbol{N}$, so that the spectral fluctuating vorticity reads
\begin{equation}
\hat{\boldsymbol{\omega}}(\boldsymbol{k},t) = k \Big[ u_+(\boldsymbol{k},t) \boldsymbol{N}(\boldsymbol{k},t) - u_- (\boldsymbol{k},t) \boldsymbol{N}^* (\boldsymbol{k},t) \Big].
\end{equation}
Considering only the positive helical modes  we get for the kinetic energy spectrum:
\begin{equation}
E_+(k,t) = \int_{S_k} \mathcal{E}_+(\boldsymbol{k},t) \mathrm{d}^2 \boldsymbol{k} 
= \int_{S_k} \langle u_+^*(\boldsymbol{k},t) u_+(\boldsymbol{k},t)\rangle \, \mathrm{d}^2 \boldsymbol{k},
\end{equation}
and the helical spectrum is thus simply given by $H_+(k,t) = k E_+(k,t)$. The evolution equation of $u_+$ is obtained by contracting \eqref{eq_NS} with $N_i^*/2$ which gives
\begin{equation}
\left(\frac{\partial}{\partial t} + \nu k^2 \right) u_+(\boldsymbol{k},t) = - \frac{1}{2} \mathrm{i} \Big(k_i N_j^*(\boldsymbol{k}) + k_j N_i^*(\boldsymbol{k}) \Big) 
\int_{\boldsymbol{k} = \boldsymbol{p} + \boldsymbol{q}} u_+(\boldsymbol{p}) u_+(\boldsymbol{q}) N_i(\boldsymbol{p}) N_j(\boldsymbol{q}) \mathrm{d}^3 \boldsymbol{p}
\end{equation}
The evolution equation of the kinetic energy density $\mathcal{E}_+$ is then
\begin{equation}
\left(\frac{\partial}{\partial t} + 2 \nu k^2 \right) \mathcal{E}_+(\boldsymbol{k},t) = T_+(\boldsymbol{k},t).
\end{equation}
It is very important to stress that the non-linear terms of the  Navier-Stokes equations conserve both energy and helicity triad-by-triad, and therefore the same is true for the dNS. In other words, the non-linear transfer $T_+$ is conservative, takes into account triadic interactions of positive helical modes, and reads after some algebra
\begin{equation}
 T_+(\boldsymbol{k},t) = \int \Re \Big[ \Big(k_i N_j^*(\boldsymbol{k}) + k_j N_i^*(\boldsymbol{k}) \Big) S_+(\boldsymbol{k},\boldsymbol{p},t) N_i(\boldsymbol{p}) N_j(\boldsymbol{q}) \Big] \mathrm{d}^3 \boldsymbol{k},
\end{equation}
where $S_+$ is the spectral triple velocity correlation
\begin{equation}
S_+(\boldsymbol{k},\boldsymbol{p},t) \delta(\boldsymbol{k} + \boldsymbol{p} + \boldsymbol{q}) = \mathrm{i} \langle u_+(\boldsymbol{k},t) u_+(\boldsymbol{p},t) u_+(\boldsymbol{q},t) \rangle.
\label{Spuuu}
\end{equation}
Then, after some technical manipulations typical of the EDQNM approximation \citep{Orszag1970,Lesieur2008},
one obtains the \textit{decimated Lin equation} of the (positive) kinetic energy spectrum, namely
\begin{equation}
\left(\frac{\partial}{\partial t} + 2 \nu k^2 \right) E_+(k,t) = S_+^E(k,t),
\label{eq_dLinE}
\end{equation}
where the spherically-averaged non-linear transfer is
\begin{align}
S_+^E(k,t) & = \int_{S_k} T_+(\boldsymbol{k},t) \mathrm{d}^2 \boldsymbol{k} = 
16 \pi^2 \int_{\Delta_k} k^2 p q \theta_{kpq} (1+z) \mathcal{E}_+'' \Bigg[p \mathcal{E}_+ (y-z)(y+z-2x) \nonumber \\
& + (1-x-2yz) \Big(k(1+y) \mathcal{E}_+' - p(1+x) \mathcal{E}_+ \Big) \Bigg] \mathrm{d}p \mathrm{d}q,
\label{transfer_dNS}
\end{align}
with $\Delta_k$ the domain where $k$, $p$ and $q$ are the lengths of the sides of the triangle formed by the triad $\boldsymbol{k} + \boldsymbol{p} + \boldsymbol{q} = \boldsymbol{0}$, and where $x$, $y$ and $z$ are the cosines of the angles formed by $\boldsymbol{p}$ and $\boldsymbol{q}$, $\boldsymbol{q}$ and $\boldsymbol{k}$, and $\boldsymbol{k}$ and $\boldsymbol{p}$ respectively. For the sake of clarity $\mathcal{E}_+=E_+(k,t)/4 \pi k^2$, $\mathcal{E}_+'=E_+(p,t)/4 \pi p^2$, and $\mathcal{E}_+''=E_+(q,t)/4 \pi q^2$. $\theta_{kpq}$ is the characteristic time of the triple correlations where the eddy-damping term is given by $A_1 \sqrt{\int_0^k u^2 E_+(u,t) \mathrm{d}u}$. Here we will start by taking
 $A_1=0.355$ for consistency with previous studies \citep{Pouquet1976, Lesieur2000, BGSM2015, BG2017}, the impact of changing $A_1$ is discussed later on.
The previous equations of the EDQNM closure for the sign-definite helicity originating from the decimated Navier-Stokes equation
are the main theoretical contributions of this work. 

For consistency and clarity, it is recalled that in isotropic helical turbulence (without decimation), the kinetic energy spectrum evolves according to $(\partial_t + 2\nu k^2)E = S_E$, where $S_E$ is the usual isotropic spherically-averaged non-linear transfer, namely
\begin{align}
S_E(k,t) = 16 \pi^2 \int_{\Delta_k} \theta_{kpq} k^2 p^2 q (xy+z^3) \mathcal{E}^{''}(\mathcal{E}^{'}-\mathcal{E}) \mathrm{d}p \mathrm{d}q,
\label{transfer_HIT}
\end{align}
where $\mathcal{E}=E(k,t)/4 \pi k^2$, $\mathcal{E}'=E(p,t)/4 \pi p^2$, and $\mathcal{E}''=E(q,t)/4 \pi q^2$. The eddy-damping term is given by the complete energy spectrum $E$, unlike in the decimated version where it is given by $E_+$. The expression \eqref{transfer_dNS} is more complicated than \eqref{transfer_HIT} in the sense that it is less compact and symmetric: this is due to the further contraction with the helical modes to select only specific triadic interactions.

In what follows, we wish to recover two features of homochiral turbulence: (i) the direct helicity cascade of \cite{Biferale2013} where the kinetic energy spectrum scales as $E_+(k) \sim {\epsilon_H}^{2/3} k^{-7/3}$  and (ii) the inverse cascade of energy in $E_+(k) \sim \epsilon^{2/3} k^{-5/3}$. For the EDQNM simulations, the wavenumber space is discretized using a logarithmic mesh $k_{i+1} = r k_i $ for $i=1, \ldots, n$, where $n$ is the total number of modes and $r=10^{1/f}$, $f=15$ being the number of points per decade. It has been checked that increasing $f$, for instance up to $f=20$, does not modify the slopes of the spectra, nor the asymptotic values of the integrated quantities (like kinetic energy) more than $1 \%$. 
This mesh spans from $k_\text{min}=10^{-6} k_L$ to $k_\text{max} = 10  k_\eta$, where $k_L$ is the integral wavenumber and $k_\eta = (\epsilon/\nu^3)^{1/4}$ the Kolmogorov wavenumber. The initial kinetic energy spectrum is given by $E_+(k) \sim k^\sigma \exp(-k^2)$, where the infrared slope is $\sigma=2$, corresponding to Saffman turbulence. Simulations not presented here revealed that the results of this work are independent of the infrared slope, in particular the findings are the same for Batchelor turbulence ($\sigma=4$). The initial $E_+$ is normalized so that $\langle u_+^2 \rangle = \int_0^\infty E_+(k) \mathrm{d}k$ is unit at $t=0$.

\section{Results at large Reynolds numbers}

\begin{figure}[!h]
	\begin{center}
	\begin{minipage}{160mm}
	\subfloat[]{
	\resizebox*{8cm}{!}{\includegraphics{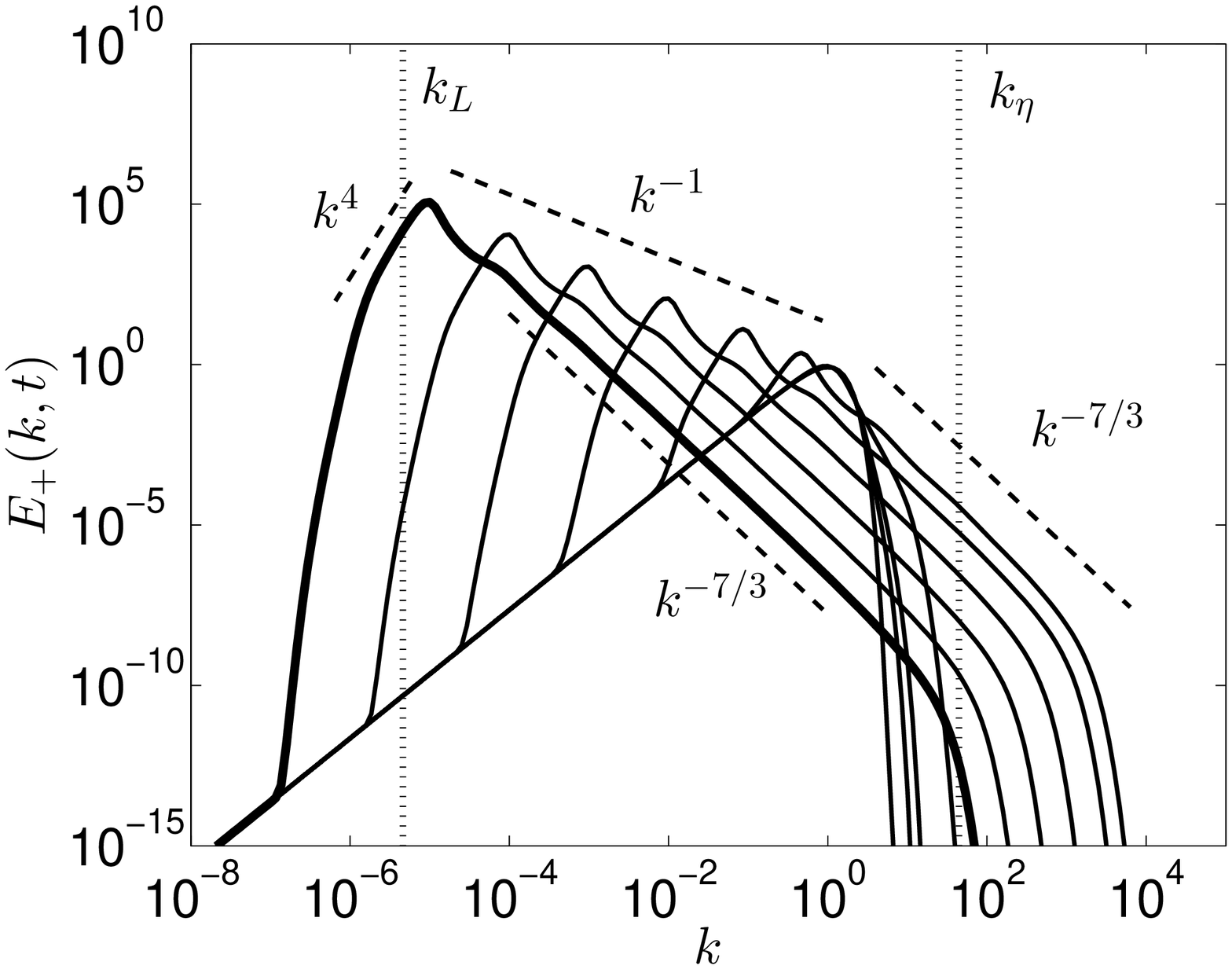}} \label{Ep_s2_time}}
	\subfloat[]{	
	\resizebox*{8cm}{!}{\includegraphics{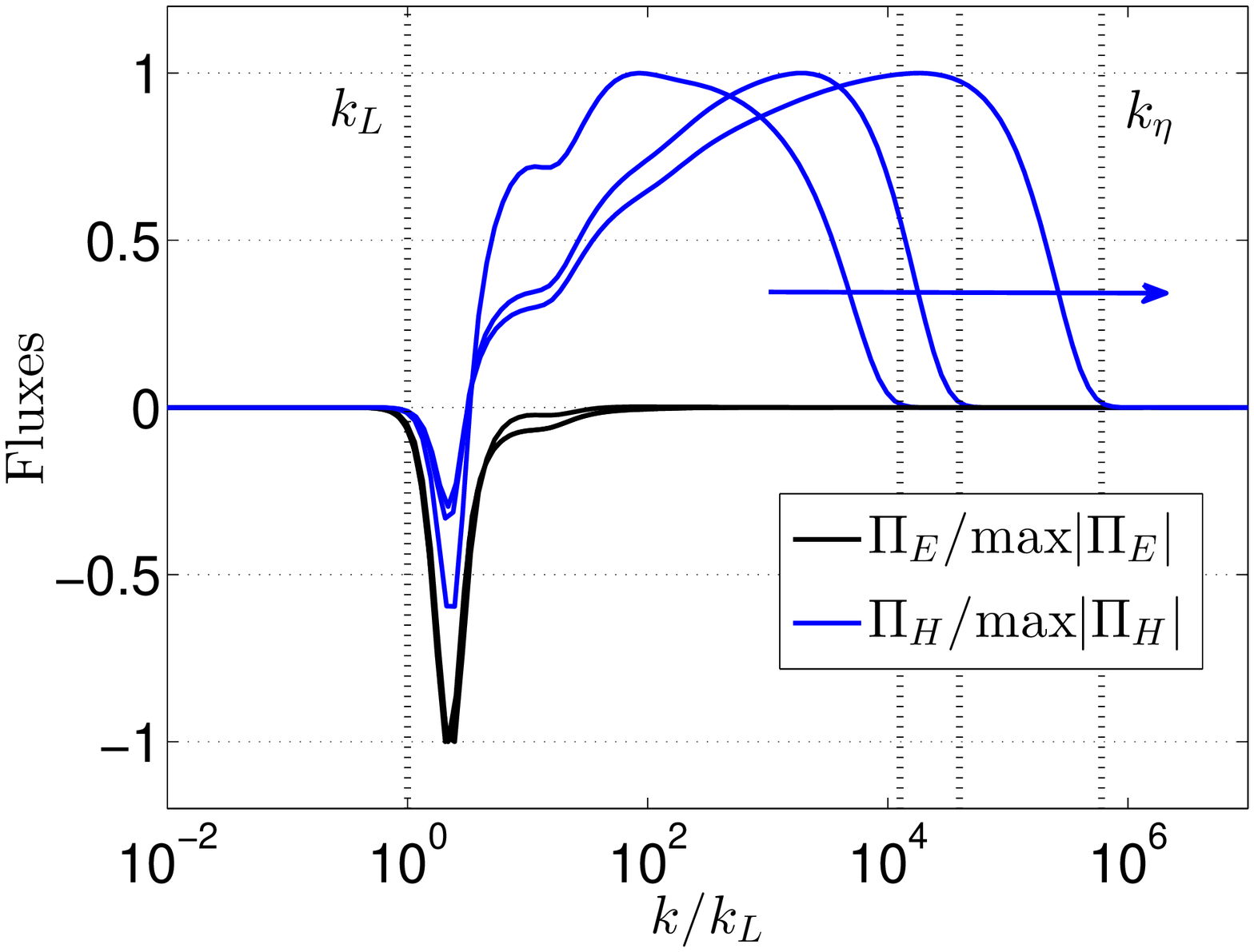}} \label{Flux_EpHp_s2}} 
 	\caption{Time evolution of the kinetic energy spectrum and of the kinetic and helical fluxes for Saffman turbulence ($\sigma=2$): the integral and Kolmogorov wavenumbers $k_L$ and $k_\eta$ are displayed as vertical dashed lines.
 	(a) $E_+(k,t)$ for various times $t=0$, $t=0.1 \tau_0$, $t=0.5 \tau_0$ and $t=10^n \tau_0$ for $n \in [1;6]$, where $\tau_0$ is the eddy turnover time: $k_L$ and $k_\eta$ correspond to the last time (thick spectrum) where $Re_\lambda(t=10^6 \tau_0)=10^6$. (b) Kinetic (black) and helical (blue) fluxes $\Pi_E$ and $\Pi_H$ for various times $t/\tau_0=[10;10^2;10^4]$. For better readability, fluxes are normalized by their maximum value and presented as functions of $k/k_L$.}
	\end{minipage}
	\end{center}
\end{figure}

First, homogeneous homochiral turbulence without any forcing is considered. The $k^{-7/3}$ scaling is recovered in figure \ref{Ep_s2_time} at large Reynolds numbers, which corresponds to the forward helicity cascade \citep{Brissaud1973}: this inertial $k^{-7/3}$ range grows with time and spans more than 4 decades at the largest Reynolds number. The noteworthy feature is that even without forcing, if large Reynolds numbers are reached, strong inverse transfers mechanisms occur since the peak of $E_+$, given at the integral wavenumber $k_L$, increases with time, unlike in fully isotropic decaying turbulence. A further evidence for intense inverse non-local transfers is that, even though the infrared scaling of the kinetic energy spectrum is initially $E_+(k<k_L) \sim k^2$, it rapidly becomes $E_+(k<k_L) \sim k^4$, which is a signature of strong back transfers \citep{Eyink2000, Lesieur2000, Meldi2012, BGSM2015}. It follows from $E_+(k) \sim k^{-7/3}$ that the helical spectrum scales in $H_+(k) = k E_+(k) \sim k^{-4/3}$.

In addition, one can remark from figure \ref{Ep_s2_time} that the peak of the kinetic energy spectrum $E_+^{\rm peak}$ seems to evolve as $k^{-1}$ with time. This has nothing to do with inertial range scaling considerations, since the inertial range scaling clearly remains $k^{-7/3}$. This can be briefly justified as follows. Since we have mainly a direct helicity cascade, one can roughly assumes that $\epsilon \simeq 0$, consistently with arguments given in \cite{Brissaud1973}. Thus, it follows that the kinetic energy $\langle u_+^2 \rangle$, whose evolution is given by $\partial_t \langle u_+^2 \rangle = - \epsilon \simeq 0$, remains constant, which is well verified numerically (see later
figure \ref{KL_s2_forcing}). Then, using the definition of the kinetic energy, one can reasonably assume  that it is mainly given by the integral scale:
\begin{equation}
\langle u_+^2 \rangle = \int_0^\infty E_+(k,t) \mathrm{d}k \simeq
\int_0^{k_L} E_+^{\rm peak}(t) \mathrm{d}k = k_L E_+^{\rm peak}(t).
\end{equation}
The kinetic energy being constant, one obtains that the time evolution of the kinetic energy spectrum peak is given by $E_+^{\rm peak}(t) \sim {k_L}^{-1}(t)$. Note that for dimensional reasons the integral scale evolves like $L \sim t$ in the unforced case. It will be shown later for the forced case that the time evolution of $\langle u_+^2 \rangle$ and $L$ are quite different. In  figure \ref{Flux_EpHp_s2} we show the evolution of the two fluxes, $\Pi_E$ and $\Pi_H$, where $\Pi_E(k) = - \int_0^k S_+^E(x) \mathrm{d}x$ and $\Pi_H(k) = - \int_0^k x S_+^E(x) \mathrm{d}x$. As said earlier, both energy and helical transfers with homochiral triadic interactions are conservative: indeed, one has $\Pi_E(k \to \infty)= \Pi_H(k \to \infty)= 0$. Furthermore, there is a direct cascade of helicity since $\Pi_H$ is mostly positive in the inertial range and spans more and more decades with time, which is consistent with the $k^{-7/3}$ scaling of $E_+$ of figure \ref{Ep_s2_time}. In addition, figure \ref{Flux_EpHp_s2} illustrates that there is an inverse cascade occurring on a small range around $k_L$ for $E_+$. \\

\begin{figure}[!h]
	\begin{center}
	\begin{minipage}{160mm}
	\subfloat[]{
	\resizebox*{8cm}{!}{\includegraphics{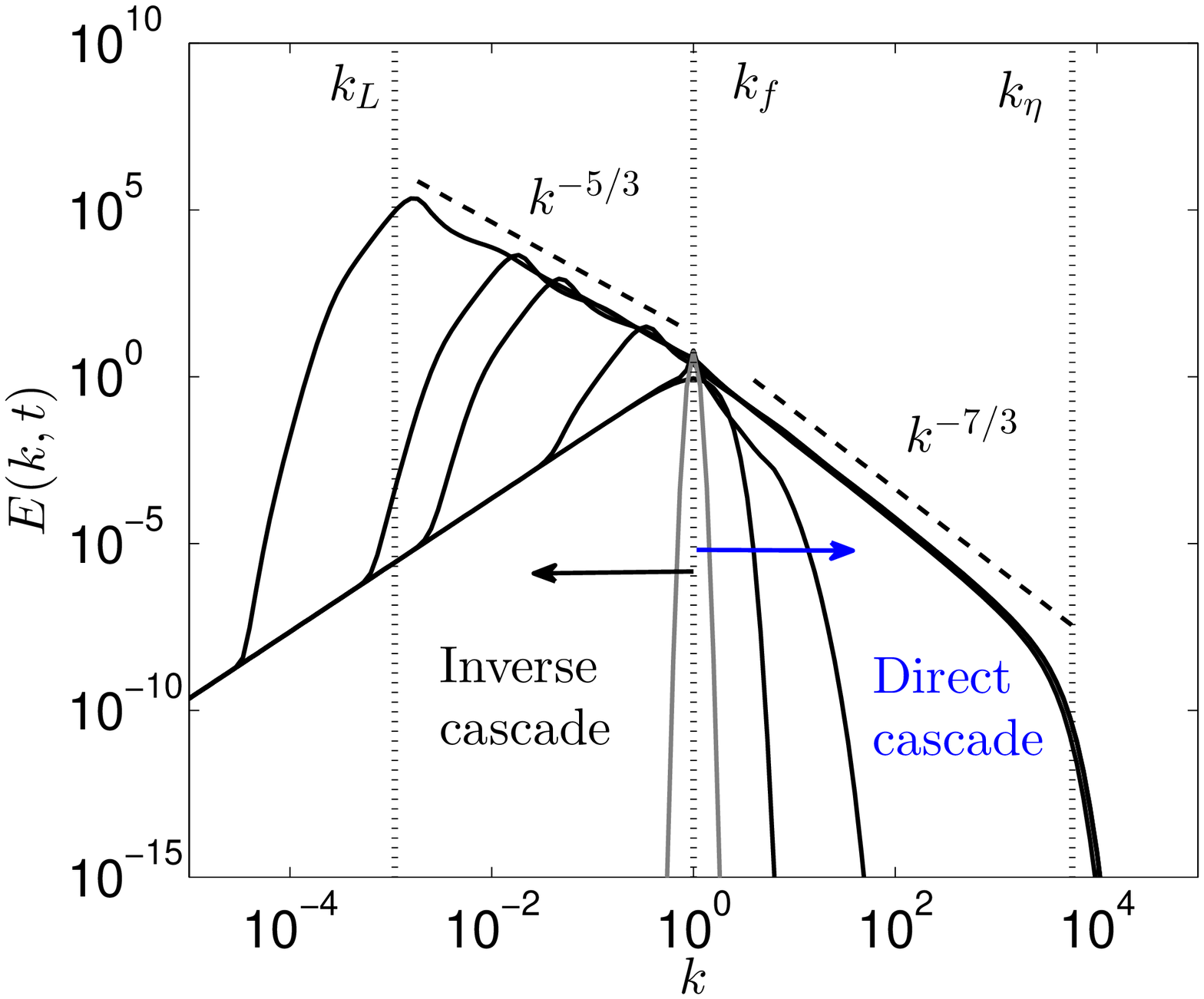}}  	\label{Ep_s2_forcing}}
	\subfloat[]{	
	\resizebox*{8cm}{!}{\includegraphics{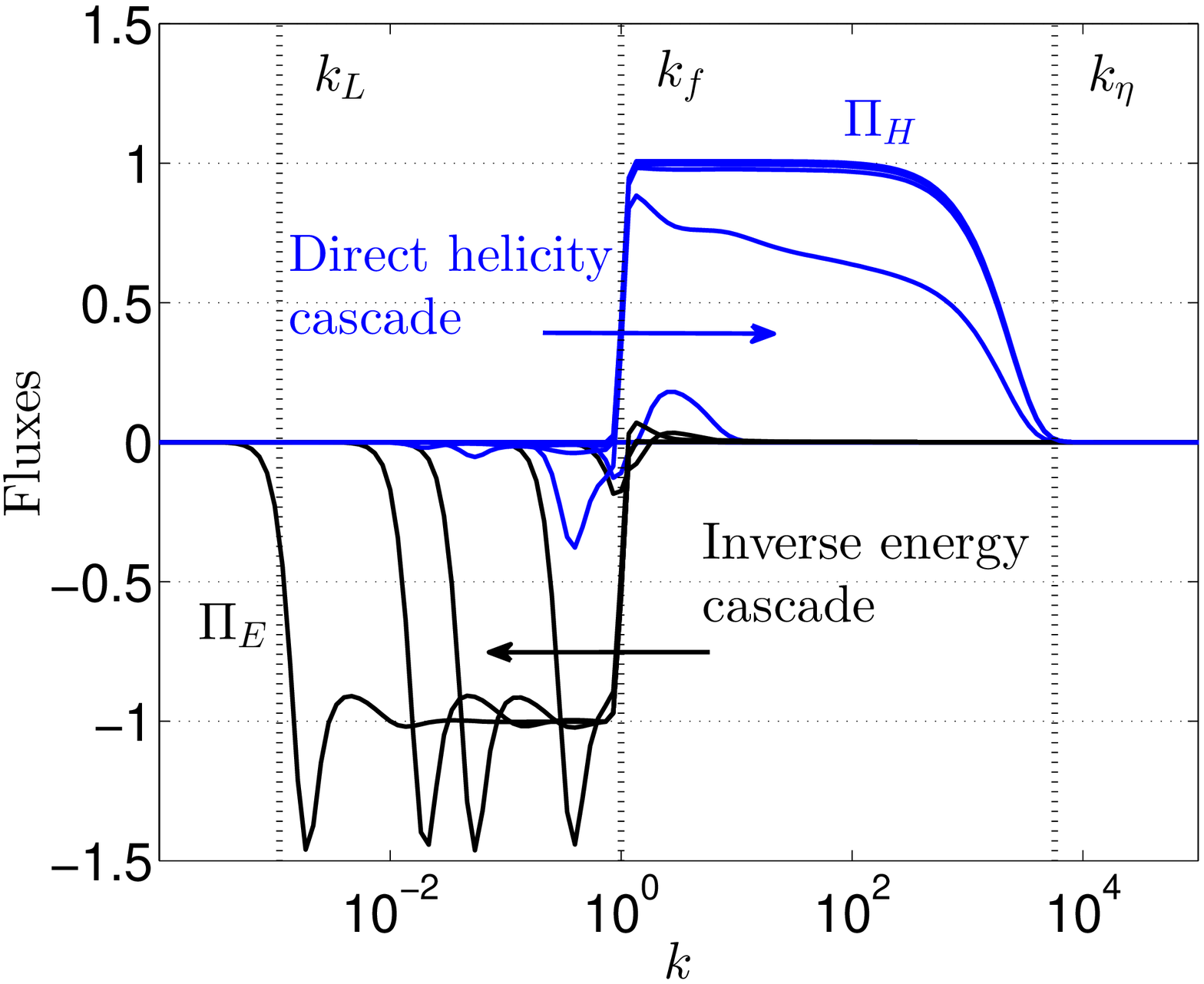}} \label{Flux_EpHp_s2_forcing}} 
 	\caption{Time evolution of the kinetic energy spectrum, kinetic and helical fluxes for Saffman turbulence ($\sigma=2$), from $t=0$ to $t=500 \tau_0$, where the curves are sampled at $t/\tau_0 =0 ; 1 ; 10 ; 50 ; 100 ; 500 $. The integral, forcing, and Kolmogorov wavenumbers $k_L$, $k_f$, and $k_\eta$  are displayed as vertical dashed lines at $Re_\lambda(t=500 \tau_0)=3.10^5$. (a) $E_+(k,t)$ with the forcing term $F(k)$ (grey) defined in \eqref{Forcing}. (b) Kinetic (black) and helical (blue) fluxes $\Pi_E$ and $\Pi_H$.}
	\end{minipage}
	\end{center}
\end{figure}

\begin{figure}[!h]
	\begin{center}
	\begin{minipage}{160mm}
	\subfloat[]{
	\resizebox*{8cm}{!}{\includegraphics{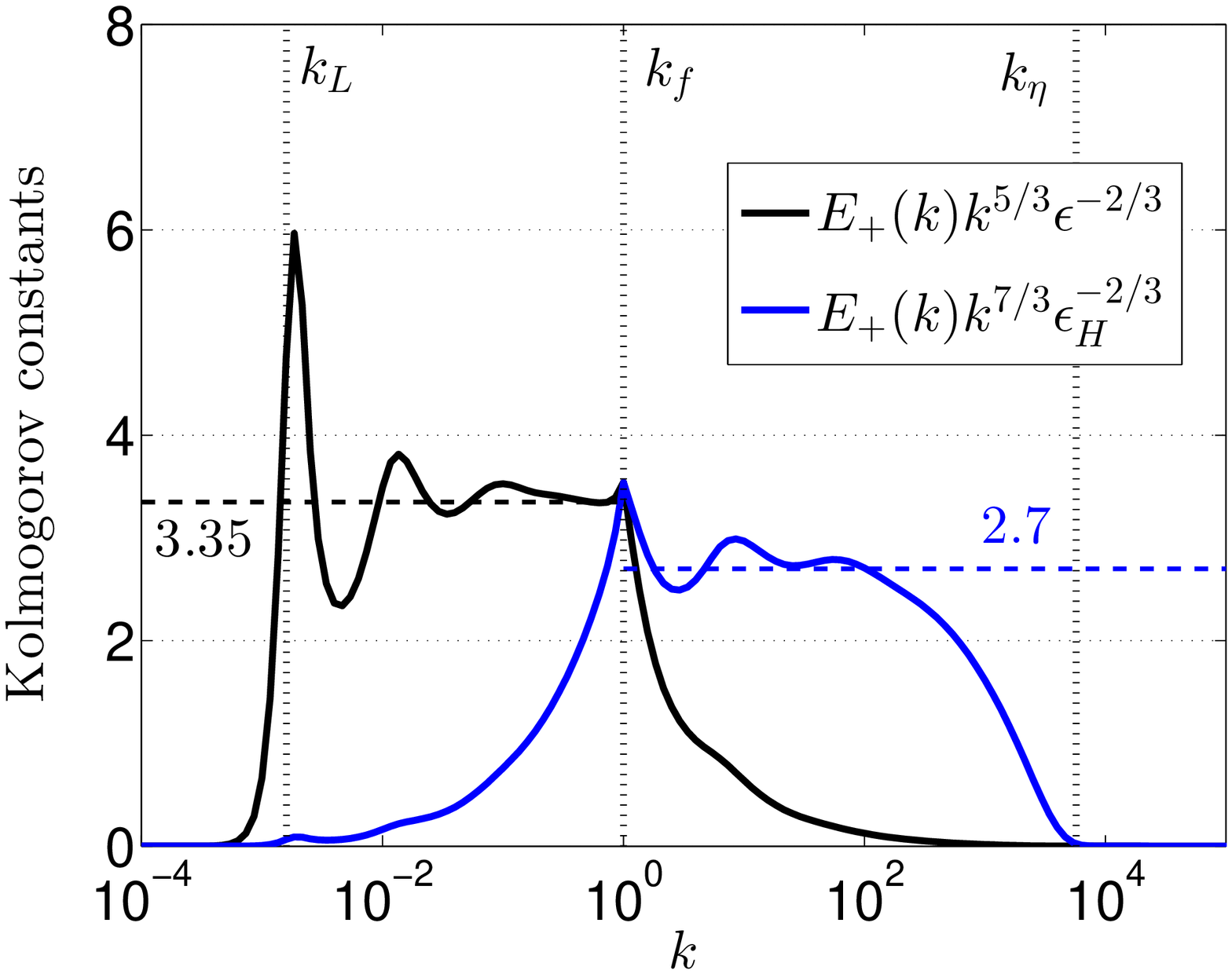} }  	
	\label{Ep_s2_comp}}
	\subfloat[]{
	\resizebox*{8cm}{!}{\includegraphics{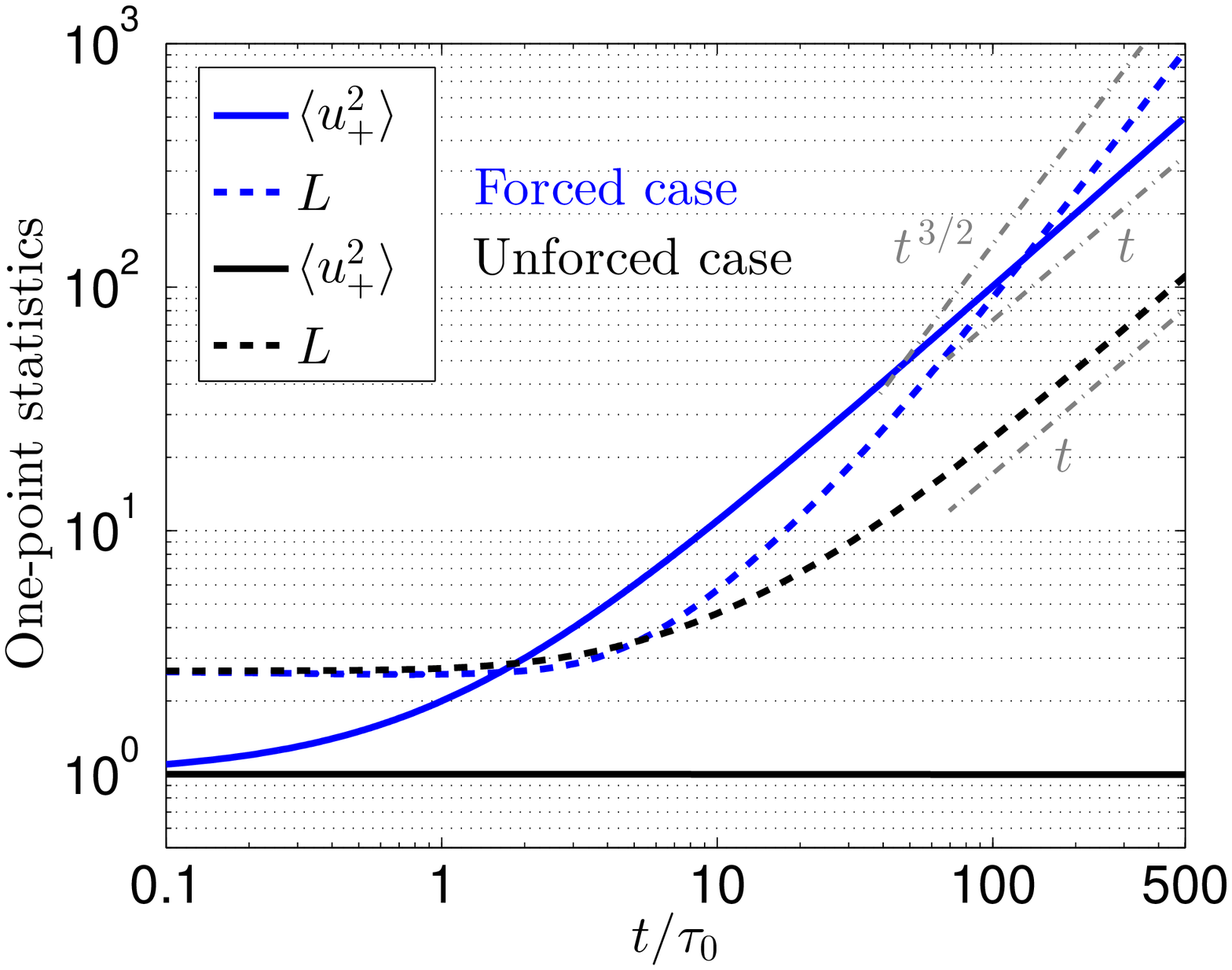} }  	
	\label{KL_s2_forcing}}
 	\caption{(a) Compensated kinetic energy spectra in the inverse (black) and direct (blue) cascades. The integral, forcing, and Kolmogorov wavenumbers $k_L$, $k_f$, and $k_\eta$  are displayed as vertical dashed lines at $Re_\lambda (t=500 \tau_0)=3.10^5$. (b) Time evolution of the kinetic energy $\langle u_+^2 \rangle$ ($-$) and integral scale $L$ ($--$) for $\sigma=2$, for the forced (blue) and unforced (black) cases. The grey $- \cdot$ curves indicate the power laws $t$ and $t^{3/2}$.}
	\end{minipage}
	\end{center}
\end{figure}

To increase the scaling region of the inverse cascade, we add to the decimated Lin equation \eqref{eq_dLinE} a forcing term $F(k)$, to have the possibility to
study the split cascade scenario with a well developed inverse energy transfer for asymptotically long times. The forcing term is given by 
\begin{equation}
F(k) = C_1 \exp\ \left(- \frac{1}{(C_2)^2} \left[ \ln \left(\frac{k}{k_f} \right) \right]^2 \right),
\label{Forcing}
\end{equation}
with $C_1$ so that one has $\int_0^\infty F(k) \mathrm{d}k=1$, $C_2=0.1$ and $k_f=1$ as proposed in \cite{BG2017}. We double checked that the following numerical results are independent of the forcing term by studying also the case with  $F(k) \sim k^4 \exp(-2k^2)$ (not shown).  
The time evolution of the kinetic energy spectrum is shown in figure \ref{Ep_s2_forcing}, where the forcing term $F$ is in grey. It is clear that the system
develops a split cascade on both sides of the forcing term. Indeed, as time increases, a $k^{-5/3}$ range grows at large scales, similar to the one obtained in \cite{Biferale2012}, which is a strong evidence for the inverse cascade of kinetic energy. Whereas for wavenumbers larger than $k_f$, the $k^{-7/3}$ range is preserved, as in figure \ref{Ep_s2_time} without forcing.  It is important to stress that this is the first time where the split energy-helicity simultaneous cascade is observed: this is notably due to the fact that by using EDQNM we can push the resolution on both sides to very large values. 
It is reasonable to conclude from figure \ref{Ep_s2_forcing} that the non-linear transfers which are at the origin of the inverse cascade are dominantly local in homochiral turbulence: indeed, because of the logarithmic discretization of the wavenumber space, elongated triads corresponding to non-local transfers cannot be taken into account with EDQNM \citep{Lesieur2008}.
This statement does not mean that there are no non-local interactions: indeed, it has been argued that some inverse non-local transfers are responsible for the change of the infrared slope of $E_+$ from $k^2$ to $k^4$ in figure \ref{Ep_s2_time} \citep{Metais1986}.

In figure \ref{Flux_EpHp_s2_forcing}  the kinetic and helical fluxes $\Pi_E$ and $\Pi_H$ are presented. 
For $k>k_f$, at scales smaller than the forcing one, the helical flux is positive, indicating a direct cascade of helicity. For $k<k_f$, the helical flux is zero and the kinetic one $\Pi_E$ is negative, showing a  stable  inverse cascade of energy for scales larger than the forcing one.
Note that the shape of $\Pi_E$ is qualitatively in agreement with the one of \cite{Sahoo2017PRL} (figure 3 therein, curve marked with squares). Here we show for the first time in a clean way that the inverse transfer has a non trivial  wavy shape around the front.
The numerical evidence for the split energy cascade in figure \ref{Flux_EpHp_s2_forcing} further justifies that for $k<k_f$ the kinetic energy depends only $\epsilon$ (since $\Pi_H$ is almost zero), and that $E_+$ and $H_+$ only depend on $\epsilon_H$ for $k>k_f$ (since $\Pi_E$ is almost zero).
%
Compensated kinetic energy spectra are presented in figure \ref{Ep_s2_comp} in the direct and inverse cascades. A reasonable plateau is obtained in both cases spanning two decades. In the inverse cascade, $E_+(k) k^{5/3} \epsilon^{-2/3}$ settles around $3.35$, and $E_+(k) k^{7/3} {\epsilon_H}^{-2/3}$ around $2.7$ in the direct cascade.
Both constants are larger than the Kolmogorov one in HIT. For the constant of the inverse cascade, this is somehow consistent with constants of inverse energy cascades found in two-dimensional turbulence which are roughly between 6 and 10 \citep{Kraichnan1971, Pouquet1975, Frisch1984}. 
Note that the values of the present constants could be modified by changing the eddy-damping parameter $A_1$, which is here chosen to be $0.355$ for consistency with previous EDQNM simulations in isotropic and skew-isotropic homogeneous turbulence \citep{BGSM2015,BG2017}. More specifically, the plateau $E_+(k) k^{5/3} \epsilon^{-2/3}$ for the inverse cascade increases with larger eddy-damping constants, namely from $2.3$ to $4.2$, for $A_1$ varying from $0.2$ to $0.49$ (this latter value for $A_1$ was used in \cite{Bos2012}). The value $4.2$ for the plateau of the inverse cascade is very close to 
what is obtained in \cite{Biferale2012}. To obtain a plateau for $E_+(k) k^{5/3} \epsilon^{-2/3}$ around 6 as in 2D turbulence, one would need to go up to $A_1=0.8$.

Finally, some one-point statistics are presented in figure \ref{KL_s2_forcing}, namely the kinetic energy $\langle u_+^2 \rangle$
and the integral scale $L= 1/k_L = 3 \pi/\langle u_+^2 \rangle \int_0^\infty (E_+(k)/k) \mathrm{d}k$ for both the forced and unforced cases.
For the unforced case, the constancy of $ \langle u_+^2 \rangle$ and $L \sim t$, discussed earlier, are recovered.
For the forced configuration, the kinetic energy is expected to grow as $ \langle u_+^2 \rangle \sim \epsilon t$  \citep{Pouquet1975}. The linear dependence with time is recovered in figure \ref{KL_s2_forcing}. Then, it follows from dimensional analysis $L \sim K^{3/2}/\epsilon$ that the integral scale should evolve like $t^{3/2}$, which is also assessed in figure \ref{KL_s2_forcing}. \\

\section{Conclusions}

In conclusion, we addressed  a particular kind of helical turbulence, where only specific triadic interactions are kept, so that the helicity is made sign-definite. In this particular homochiral framework, the 3D turbulence possesses two sign-definite inviscid invariants, namely kinetic energy and helicity, like 2D turbulence with kinetic energy and enstrophy. 
The main objective of this study was to show that a spectral closure method, such as EDQNM, could recover the main findings of \cite{Biferale2012,Biferale2013}. To do so, an adapted EDQNM approximation was derived, taking into account only the particular interactions that make helicity sign-definite (positive here).
The direct helicity cascade, where $E_+(k) \sim {\epsilon_H}^{2/3} k^{-7/3}$, is first obtained in unforced homochiral turbulence, where the kinetic energy flux $\epsilon \simeq 0$. In such a configuration, the kinetic energy $\langle u_+^2 \rangle$ remains constant and the peak of $E_+$  evolves in ${k_L}^{-1}$ with time.
When a forcing term is added, in addition to the helicity forward cascade, an inverse energy cascade  develops toward smaller wavenumbers, where the helicity flux $\epsilon_H \simeq 0$. In this regime of forced homochiral turbulence, the kinetic energy evolves linearly with time $\langle u_+^2 \rangle \sim \epsilon t$, and the integral scale like $L \sim t^{3/2}$. 

Our work shows for the first time the possibility to have a stable split cascade in three dimensional turbulence under strong restriction of the helical interactions. It remains to be understood how much this phenomenology might be indeed observed in more realistic flow conﬁgurations, as in the presence of strong rotation or confinement, where also a transition from direct to inverse energy cascade is observed \citep{Mininni2009, Celani2010, Xia2011, Godeferd2015, Biferale2016}. It is also interesting to study the inverse cascade regime in the presence of a large scale drag, in order to allow for the formation of a large scale stationary condensate \citep{Chertkov2007}.  At difference from the two dimensional case, the condensate will necessarily have strong helical properties and be close to an ABC quasi-stationary solution of the three dimensional Navier-Stokes equations \citep{Dombre1986, Moffatt2014, Sahoo2015bis}. Work in this direction will be reported elsewhere. Finally, it is important to stress that similar EDQNM studies can be extended to helical MHD \citep{Linkmann2017}.

The research leading to these results has received funding from the European
Union's Seventh Framework Programme (FP7/2007-2013) under grant agreement No.
339032.



\bibliography{HHT_BBG}

\providecommand{\noopsort}[1]{}\providecommand{\singleletter}[1]{#1}%
\begin{thebibliography}{38}%
\makeatletter
\providecommand \@ifxundefined [1]{%
 \@ifx{#1\undefined}
}%
\providecommand \@ifnum [1]{%
 \ifnum #1\expandafter \@firstoftwo
 \else \expandafter \@secondoftwo
 \fi
}%
\providecommand \@ifx [1]{%
 \ifx #1\expandafter \@firstoftwo
 \else \expandafter \@secondoftwo
 \fi
}%
\providecommand \natexlab [1]{#1}%
\providecommand \enquote  [1]{``#1''}%
\providecommand \bibnamefont  [1]{#1}%
\providecommand \bibfnamefont [1]{#1}%
\providecommand \citenamefont [1]{#1}%
\providecommand \href@noop [0]{\@secondoftwo}%
\providecommand \href [0]{\begingroup \@sanitize@url \@href}%
\providecommand \@href[1]{\@@startlink{#1}\@@href}%
\providecommand \@@href[1]{\endgroup#1\@@endlink}%
\providecommand \@sanitize@url [0]{\catcode `\\12\catcode `\$12\catcode
  `\&12\catcode `\#12\catcode `\^12\catcode `\_12\catcode `\%12\relax}%
\providecommand \@@startlink[1]{}%
\providecommand \@@endlink[0]{}%
\providecommand \url  [0]{\begingroup\@sanitize@url \@url }%
\providecommand \@url [1]{\endgroup\@href {#1}{\urlprefix }}%
\providecommand \urlprefix  [0]{URL }%
\providecommand \Eprint [0]{\href }%
\providecommand \doibase [0]{http://dx.doi.org/}%
\providecommand \selectlanguage [0]{\@gobble}%
\providecommand \bibinfo  [0]{\@secondoftwo}%
\providecommand \bibfield  [0]{\@secondoftwo}%
\providecommand \translation [1]{[#1]}%
\providecommand \BibitemOpen [0]{}%
\providecommand \bibitemStop [0]{}%
\providecommand \bibitemNoStop [0]{.\EOS\space}%
\providecommand \EOS [0]{\spacefactor3000\relax}%
\providecommand \BibitemShut  [1]{\csname bibitem#1\endcsname}%
\let\auto@bib@innerbib\@empty
\bibitem [{\citenamefont {Moffatt}(1969)}]{Moffatt1969}%
  \BibitemOpen
  \bibfield  {author} {\bibinfo {author} {\bibfnamefont {H.~K.}\ \bibnamefont
  {Moffatt}},\ }\bibfield  {title} {\enquote {\bibinfo {title} {The degree of
  knottedness of tangled vortex lines},}\ }\href@noop {} {\bibfield  {journal}
  {\bibinfo  {journal} {Journal of Fluid Mechanics}\ }\textbf {\bibinfo
  {volume} {35}},\ \bibinfo {pages} {117--129} (\bibinfo {year}
  {1969})}\BibitemShut {NoStop}%
\bibitem [{\citenamefont {Boffetta}\ and\ \citenamefont
  {Ecke}(2012)}]{Boffetta2012}%
  \BibitemOpen
  \bibfield  {author} {\bibinfo {author} {\bibfnamefont {G.}~\bibnamefont
  {Boffetta}}\ and\ \bibinfo {author} {\bibfnamefont {R.~E.}\ \bibnamefont
  {Ecke}},\ }\bibfield  {title} {\enquote {\bibinfo {title} {Two-dimensional
  turbulence},}\ }\href@noop {} {\bibfield  {journal} {\bibinfo  {journal}
  {Annual Review of Fluid Mechanics}\ }\textbf {\bibinfo {volume} {44}},\
  \bibinfo {pages} {427--451} (\bibinfo {year} {2012})}\BibitemShut {NoStop}%
\bibitem [{\citenamefont {Brissaud}\ \emph {et~al.}(1973)\citenamefont
  {Brissaud}, \citenamefont {Frisch}, \citenamefont {Leorat}, \citenamefont
  {Lesieur},\ and\ \citenamefont {Mazure}}]{Brissaud1973}%
  \BibitemOpen
  \bibfield  {author} {\bibinfo {author} {\bibfnamefont {A.}~\bibnamefont
  {Brissaud}}, \bibinfo {author} {\bibfnamefont {U.}~\bibnamefont {Frisch}},
  \bibinfo {author} {\bibfnamefont {J.}~\bibnamefont {Leorat}}, \bibinfo
  {author} {\bibfnamefont {M.}~\bibnamefont {Lesieur}}, \ and\ \bibinfo
  {author} {\bibfnamefont {A.}~\bibnamefont {Mazure}},\ }\bibfield  {title}
  {\enquote {\bibinfo {title} {Helicity cascades in fully developed isotropic
  turbulence},}\ }\href@noop {} {\bibfield  {journal} {\bibinfo  {journal} {The
  Physics of Fluids}\ }\textbf {\bibinfo {volume} {16}},\ \bibinfo {pages}
  {1366--1367} (\bibinfo {year} {1973})}\BibitemShut {NoStop}%
\bibitem [{\citenamefont {Andr\'e}\ and\ \citenamefont
  {Lesieur}(1977)}]{Andre1977}%
  \BibitemOpen
  \bibfield  {author} {\bibinfo {author} {\bibfnamefont {J.~C.}\ \bibnamefont
  {Andr\'e}}\ and\ \bibinfo {author} {\bibfnamefont {M.}~\bibnamefont
  {Lesieur}},\ }\bibfield  {title} {\enquote {\bibinfo {title} {Influence of
  helicity on the evolution of isotropic turbulence at high reynolds number},}\
  }\href@noop {} {\bibfield  {journal} {\bibinfo  {journal} {Journal of Fluid
  Mechanics}\ }\textbf {\bibinfo {volume} {81}},\ \bibinfo {pages} {187--207}
  (\bibinfo {year} {1977})}\BibitemShut {NoStop}%
\bibitem [{\citenamefont {Orszag}(1970)}]{Orszag1970}%
  \BibitemOpen
  \bibfield  {author} {\bibinfo {author} {\bibfnamefont {S.~A.}\ \bibnamefont
  {Orszag}},\ }\bibfield  {title} {\enquote {\bibinfo {title} {Analytical
  theories of turbulence},}\ }\href@noop {} {\bibfield  {journal} {\bibinfo
  {journal} {Journal of Fluid Mechanics}\ }\textbf {\bibinfo {volume} {41}},\
  \bibinfo {pages} {363--386} (\bibinfo {year} {1970})}\BibitemShut {NoStop}%
\bibitem [{\citenamefont {Lesieur}(2008)}]{Lesieur2008}%
  \BibitemOpen
  \bibfield  {author} {\bibinfo {author} {\bibfnamefont {M.}~\bibnamefont
  {Lesieur}},\ }\href@noop {} {\emph {\bibinfo {title} {Turbulence in
  fluids}}}\ (\bibinfo  {publisher} {Springer, 4th Edition},\ \bibinfo
  {address} {Dordrecht},\ \bibinfo {year} {2008})\BibitemShut {NoStop}%
\bibitem [{\citenamefont {Sagaut}\ and\ \citenamefont
  {Cambon}(2008)}]{Sagaut2008}%
  \BibitemOpen
  \bibfield  {author} {\bibinfo {author} {\bibfnamefont {P.}~\bibnamefont
  {Sagaut}}\ and\ \bibinfo {author} {\bibfnamefont {C.}~\bibnamefont
  {Cambon}},\ }\href@noop {} {\emph {\bibinfo {title} {Homogeneous Turbulence
  Dynamics}}}\ (\bibinfo  {publisher} {Cambridge University Press},\ \bibinfo
  {year} {2008})\BibitemShut {NoStop}%
\bibitem [{\citenamefont {Borue}\ and\ \citenamefont
  {Orszag}(1997)}]{Borue1997}%
  \BibitemOpen
  \bibfield  {author} {\bibinfo {author} {\bibfnamefont {V.}~\bibnamefont
  {Borue}}\ and\ \bibinfo {author} {\bibfnamefont {S.A.}\ \bibnamefont
  {Orszag}},\ }\bibfield  {title} {\enquote {\bibinfo {title} {Spectra in
  helical three-dimensional homogeneous isotropic turbulence},}\ }\href@noop {}
  {\bibfield  {journal} {\bibinfo  {journal} {Physical Review E}\ }\textbf
  {\bibinfo {volume} {55}},\ \bibinfo {pages} {7005--7009} (\bibinfo {year}
  {1997})}\BibitemShut {NoStop}%
\bibitem [{\citenamefont {Chen}\ \emph {et~al.}(2003)\citenamefont {Chen},
  \citenamefont {Chen},\ and\ \citenamefont {Eyink}}]{Chen2003}%
  \BibitemOpen
  \bibfield  {author} {\bibinfo {author} {\bibfnamefont {Q.}~\bibnamefont
  {Chen}}, \bibinfo {author} {\bibfnamefont {S.}~\bibnamefont {Chen}}, \ and\
  \bibinfo {author} {\bibfnamefont {G.~L.}\ \bibnamefont {Eyink}},\ }\bibfield
  {title} {\enquote {\bibinfo {title} {The joint cascade of energy and helicity
  in three-dimensional turbulence},}\ }\href@noop {} {\bibfield  {journal}
  {\bibinfo  {journal} {Physics of Fluids}\ }\textbf {\bibinfo {volume} {15}},\
  \bibinfo {pages} {361} (\bibinfo {year} {2003})}\BibitemShut {NoStop}%
\bibitem [{\citenamefont {Briard}\ and\ \citenamefont {Gomez}(2017)}]{BG2017}%
  \BibitemOpen
  \bibfield  {author} {\bibinfo {author} {\bibfnamefont {A.}~\bibnamefont
  {Briard}}\ and\ \bibinfo {author} {\bibfnamefont {T.}~\bibnamefont {Gomez}},\
  }\bibfield  {title} {\enquote {\bibinfo {title} {Dynamics of helicity in
  homogeneous skew-isotropic turbulence},}\ }\href@noop {} {\bibfield
  {journal} {\bibinfo  {journal} {Journal of Fluid Mechanics}\ }\textbf
  {\bibinfo {volume} {821}},\ \bibinfo {pages} {539--581} (\bibinfo {year}
  {2017})}\BibitemShut {NoStop}%
\bibitem [{\citenamefont {Biferale}\ \emph {et~al.}(2012)\citenamefont
  {Biferale}, \citenamefont {Musacchio},\ and\ \citenamefont
  {Toschi}}]{Biferale2012}%
  \BibitemOpen
  \bibfield  {author} {\bibinfo {author} {\bibfnamefont {L.}~\bibnamefont
  {Biferale}}, \bibinfo {author} {\bibfnamefont {S.}~\bibnamefont {Musacchio}},
  \ and\ \bibinfo {author} {\bibfnamefont {F.}~\bibnamefont {Toschi}},\
  }\bibfield  {title} {\enquote {\bibinfo {title} {Inverse energy cascade in
  three-dimensional isotropic turbulence},}\ }\href@noop {} {\bibfield
  {journal} {\bibinfo  {journal} {Physical Review Letters}\ }\textbf {\bibinfo
  {volume} {108}},\ \bibinfo {pages} {164501} (\bibinfo {year}
  {2012})}\BibitemShut {NoStop}%
\bibitem [{\citenamefont {Biferale}\ \emph {et~al.}(2013)\citenamefont
  {Biferale}, \citenamefont {Musacchio},\ and\ \citenamefont
  {Toschi}}]{Biferale2013}%
  \BibitemOpen
  \bibfield  {author} {\bibinfo {author} {\bibfnamefont {L.}~\bibnamefont
  {Biferale}}, \bibinfo {author} {\bibfnamefont {S.}~\bibnamefont {Musacchio}},
  \ and\ \bibinfo {author} {\bibfnamefont {F.}~\bibnamefont {Toschi}},\
  }\bibfield  {title} {\enquote {\bibinfo {title} {Split energy–helicity
  cascades in three-dimensional homogeneous and isotropic turbulence},}\
  }\href@noop {} {\bibfield  {journal} {\bibinfo  {journal} {Journal of Fluid
  Mechanics}\ }\textbf {\bibinfo {volume} {730}},\ \bibinfo {pages} {309--327}
  (\bibinfo {year} {2013})}\BibitemShut {NoStop}%
\bibitem [{\citenamefont {Waleffe}(1992)}]{Waleffe1992}%
  \BibitemOpen
  \bibfield  {author} {\bibinfo {author} {\bibfnamefont {F.}~\bibnamefont
  {Waleffe}},\ }\bibfield  {title} {\enquote {\bibinfo {title} {The nature of
  triad interactions in homogeneous turbulence},}\ }\href@noop {} {\bibfield
  {journal} {\bibinfo  {journal} {Physics of Fluids}\ }\textbf {\bibinfo
  {volume} {4}},\ \bibinfo {pages} {350--363} (\bibinfo {year}
  {1992})}\BibitemShut {NoStop}%
\bibitem [{\citenamefont {Sahoo}\ \emph {et~al.}(2015)\citenamefont {Sahoo},
  \citenamefont {Bonaccorso},\ and\ \citenamefont {Biferale}}]{Sahoo2015}%
  \BibitemOpen
  \bibfield  {author} {\bibinfo {author} {\bibfnamefont {G.}~\bibnamefont
  {Sahoo}}, \bibinfo {author} {\bibfnamefont {F.}~\bibnamefont {Bonaccorso}}, \
  and\ \bibinfo {author} {\bibfnamefont {L.}~\bibnamefont {Biferale}},\
  }\bibfield  {title} {\enquote {\bibinfo {title} {Role of helicity for large-
  and small-scale turbulent fluctuations},}\ }\href@noop {} {\bibfield
  {journal} {\bibinfo  {journal} {Physical Review E}\ }\textbf {\bibinfo
  {volume} {92}},\ \bibinfo {pages} {051002(R)} (\bibinfo {year}
  {2015})}\BibitemShut {NoStop}%
\bibitem [{\citenamefont {Sahoo}\ \emph
  {et~al.}(2017{\natexlab{a}})\citenamefont {Sahoo}, \citenamefont {Alexakis},\
  and\ \citenamefont {Biferale}}]{Sahoo2017PRL}%
  \BibitemOpen
  \bibfield  {author} {\bibinfo {author} {\bibfnamefont {G.}~\bibnamefont
  {Sahoo}}, \bibinfo {author} {\bibfnamefont {A.}~\bibnamefont {Alexakis}}, \
  and\ \bibinfo {author} {\bibfnamefont {L.}~\bibnamefont {Biferale}},\
  }\bibfield  {title} {\enquote {\bibinfo {title} {Discontinuous transition
  from direct to inverse cascade in three-dimensional turbulence},}\
  }\href@noop {} {\bibfield  {journal} {\bibinfo  {journal} {Physical Review
  Letters}\ }\textbf {\bibinfo {volume} {118}},\ \bibinfo {pages} {164501}
  (\bibinfo {year} {2017}{\natexlab{a}})}\BibitemShut {NoStop}%
\bibitem [{\citenamefont {Sahoo}\ \emph
  {et~al.}(2017{\natexlab{b}})\citenamefont {Sahoo}, \citenamefont {Pietro},\
  and\ \citenamefont {Biferale}}]{Sahoo2017PRF}%
  \BibitemOpen
  \bibfield  {author} {\bibinfo {author} {\bibfnamefont {G.}~\bibnamefont
  {Sahoo}}, \bibinfo {author} {\bibfnamefont {M.~De}\ \bibnamefont {Pietro}}, \
  and\ \bibinfo {author} {\bibfnamefont {L.}~\bibnamefont {Biferale}},\
  }\bibfield  {title} {\enquote {\bibinfo {title} {Helicity statistics in
  homogeneous and isotropic turbulence and turbulence models},}\ }\href@noop {}
  {\bibfield  {journal} {\bibinfo  {journal} {Physical Review Fluids}\ }\textbf
  {\bibinfo {volume} {2}},\ \bibinfo {pages} {024601} (\bibinfo {year}
  {2017}{\natexlab{b}})}\BibitemShut {NoStop}%
\bibitem [{\citenamefont {Pouquet}\ \emph {et~al.}(1975)\citenamefont
  {Pouquet}, \citenamefont {Lesieur}, \citenamefont {Andr\'e},\ and\
  \citenamefont {Basdevant}}]{Pouquet1975}%
  \BibitemOpen
  \bibfield  {author} {\bibinfo {author} {\bibfnamefont {A.}~\bibnamefont
  {Pouquet}}, \bibinfo {author} {\bibfnamefont {M.}~\bibnamefont {Lesieur}},
  \bibinfo {author} {\bibfnamefont {J.~C.}\ \bibnamefont {Andr\'e}}, \ and\
  \bibinfo {author} {\bibfnamefont {C.}~\bibnamefont {Basdevant}},\ }\bibfield
  {title} {\enquote {\bibinfo {title} {Evolution of high reynolds number
  two-dimensional turbulence},}\ }\href@noop {} {\bibfield  {journal} {\bibinfo
   {journal} {Journal of Fluid Mechanics}\ }\textbf {\bibinfo {volume} {72}},\
  \bibinfo {pages} {305--319} (\bibinfo {year} {1975})}\BibitemShut {NoStop}%
\bibitem [{\citenamefont {Pouquet}\ \emph {et~al.}(1976)\citenamefont
  {Pouquet}, \citenamefont {Frisch},\ and\ \citenamefont
  {L\'eorat}}]{Pouquet1976}%
  \BibitemOpen
  \bibfield  {author} {\bibinfo {author} {\bibfnamefont {A.}~\bibnamefont
  {Pouquet}}, \bibinfo {author} {\bibfnamefont {U.}~\bibnamefont {Frisch}}, \
  and\ \bibinfo {author} {\bibfnamefont {J.-L.}\ \bibnamefont {L\'eorat}},\
  }\bibfield  {title} {\enquote {\bibinfo {title} {Strong mhd helical
  turbulence and the nonlinear dynamo effect},}\ }\href@noop {} {\bibfield
  {journal} {\bibinfo  {journal} {Journal of Fluid Mechanics}\ }\textbf
  {\bibinfo {volume} {77}},\ \bibinfo {pages} {321--354} (\bibinfo {year}
  {1976})}\BibitemShut {NoStop}%
\bibitem [{\citenamefont {Cambon}\ and\ \citenamefont
  {Jacquin}(1989)}]{Cambon1989}%
  \BibitemOpen
  \bibfield  {author} {\bibinfo {author} {\bibfnamefont {C.}~\bibnamefont
  {Cambon}}\ and\ \bibinfo {author} {\bibfnamefont {L.}~\bibnamefont
  {Jacquin}},\ }\bibfield  {title} {\enquote {\bibinfo {title} {Spectral
  approach to non-isotropic turbulence subjected to rotation},}\ }\href@noop {}
  {\bibfield  {journal} {\bibinfo  {journal} {Journal of Fluid Mechanics}\
  }\textbf {\bibinfo {volume} {202}},\ \bibinfo {pages} {295--317} (\bibinfo
  {year} {1989})}\BibitemShut {NoStop}%
\bibitem [{\citenamefont {von K\'arm\'an}\ and\ \citenamefont
  {Lin}(1949)}]{Lin1949}%
  \BibitemOpen
  \bibfield  {author} {\bibinfo {author} {\bibfnamefont {T.}~\bibnamefont {von
  K\'arm\'an}}\ and\ \bibinfo {author} {\bibfnamefont {C.~C.}\ \bibnamefont
  {Lin}},\ }\bibfield  {title} {\enquote {\bibinfo {title} {On the concept of
  similiarity in the theory of isotropic turbulence},}\ }\href@noop {}
  {\bibfield  {journal} {\bibinfo  {journal} {Rev. Mod. Phys.}\ }\textbf
  {\bibinfo {volume} {21}},\ \bibinfo {pages} {516--519} (\bibinfo {year}
  {1949})}\BibitemShut {NoStop}%
\bibitem [{\citenamefont {Lesieur}\ and\ \citenamefont
  {Ossia}(2000)}]{Lesieur2000}%
  \BibitemOpen
  \bibfield  {author} {\bibinfo {author} {\bibfnamefont {M.}~\bibnamefont
  {Lesieur}}\ and\ \bibinfo {author} {\bibfnamefont {S.}~\bibnamefont
  {Ossia}},\ }\bibfield  {title} {\enquote {\bibinfo {title} {3d isotropic
  turbulence at very high reynolds numbers: Edqnm study},}\ }\href@noop {}
  {\bibfield  {journal} {\bibinfo  {journal} {Journal of Turbulence}\ }\textbf
  {\bibinfo {volume} {1}} (\bibinfo {year} {2000})}\BibitemShut {NoStop}%
\bibitem [{\citenamefont {Briard}\ \emph {et~al.}(2015)\citenamefont {Briard},
  \citenamefont {Gomez}, \citenamefont {Sagaut},\ and\ \citenamefont
  {Memari}}]{BGSM2015}%
  \BibitemOpen
  \bibfield  {author} {\bibinfo {author} {\bibfnamefont {A.}~\bibnamefont
  {Briard}}, \bibinfo {author} {\bibfnamefont {T.}~\bibnamefont {Gomez}},
  \bibinfo {author} {\bibfnamefont {P.}~\bibnamefont {Sagaut}}, \ and\ \bibinfo
  {author} {\bibfnamefont {S.}~\bibnamefont {Memari}},\ }\bibfield  {title}
  {\enquote {\bibinfo {title} {Passive scalar decay laws in isotropic
  turbulence: Prandtl effects},}\ }\href@noop {} {\bibfield  {journal}
  {\bibinfo  {journal} {Journal of Fluid Mechanics}\ }\textbf {\bibinfo
  {volume} {784}},\ \bibinfo {pages} {274--303} (\bibinfo {year}
  {2015})}\BibitemShut {NoStop}%
\bibitem [{\citenamefont {Eyink}\ and\ \citenamefont
  {Thomson}(2000)}]{Eyink2000}%
  \BibitemOpen
  \bibfield  {author} {\bibinfo {author} {\bibfnamefont {G.~L.}\ \bibnamefont
  {Eyink}}\ and\ \bibinfo {author} {\bibfnamefont {D.~J.}\ \bibnamefont
  {Thomson}},\ }\bibfield  {title} {\enquote {\bibinfo {title} {Free decay of
  turbulence and breakdown of self-similarity},}\ }\href@noop {} {\bibfield
  {journal} {\bibinfo  {journal} {Physics of Fluids}\ }\textbf {\bibinfo
  {volume} {12}},\ \bibinfo {pages} {477--479} (\bibinfo {year}
  {2000})}\BibitemShut {NoStop}%
\bibitem [{\citenamefont {Meldi}\ and\ \citenamefont
  {Sagaut}(2012)}]{Meldi2012}%
  \BibitemOpen
  \bibfield  {author} {\bibinfo {author} {\bibfnamefont {M.}~\bibnamefont
  {Meldi}}\ and\ \bibinfo {author} {\bibfnamefont {P.}~\bibnamefont {Sagaut}},\
  }\bibfield  {title} {\enquote {\bibinfo {title} {On non-self-similar regimes
  in homogeneous isotropic turbulence decay},}\ }\href@noop {} {\bibfield
  {journal} {\bibinfo  {journal} {Journal of Fluid Mechanics}\ }\textbf
  {\bibinfo {volume} {711}},\ \bibinfo {pages} {364--393} (\bibinfo {year}
  {2012})}\BibitemShut {NoStop}%
\bibitem [{\citenamefont {M\'etais}\ and\ \citenamefont
  {Lesieur}(1986)}]{Metais1986}%
  \BibitemOpen
  \bibfield  {author} {\bibinfo {author} {\bibfnamefont {O.}~\bibnamefont
  {M\'etais}}\ and\ \bibinfo {author} {\bibfnamefont {M.}~\bibnamefont
  {Lesieur}},\ }\bibfield  {title} {\enquote {\bibinfo {title} {Statistical
  predictability of decaying turbulence},}\ }\href@noop {} {\bibfield
  {journal} {\bibinfo  {journal} {Journal of the atmospheric sciences}\
  }\textbf {\bibinfo {volume} {43}},\ \bibinfo {pages} {857--870} (\bibinfo
  {year} {1986})}\BibitemShut {NoStop}%
\bibitem [{\citenamefont {Kraichnan}(1971)}]{Kraichnan1971}%
  \BibitemOpen
  \bibfield  {author} {\bibinfo {author} {\bibfnamefont {R.~H.}\ \bibnamefont
  {Kraichnan}},\ }\bibfield  {title} {\enquote {\bibinfo {title}
  {Inertial-range transfer in two- and three-dimensional turbulence},}\
  }\href@noop {} {\bibfield  {journal} {\bibinfo  {journal} {Journal of Fluid
  Mechanics}\ }\textbf {\bibinfo {volume} {47}},\ \bibinfo {pages} {525--535}
  (\bibinfo {year} {1971})}\BibitemShut {NoStop}%
\bibitem [{\citenamefont {Frisch}\ and\ \citenamefont
  {Sulem}(1984)}]{Frisch1984}%
  \BibitemOpen
  \bibfield  {author} {\bibinfo {author} {\bibfnamefont {U.}~\bibnamefont
  {Frisch}}\ and\ \bibinfo {author} {\bibfnamefont {P.~L.}\ \bibnamefont
  {Sulem}},\ }\bibfield  {title} {\enquote {\bibinfo {title} {Numerical
  simulation of the inverse cascade in two-dimensional turbulence},}\
  }\href@noop {} {\bibfield  {journal} {\bibinfo  {journal} {Physics of
  Fluids}\ }\textbf {\bibinfo {volume} {27}},\ \bibinfo {pages} {1921--1923}
  (\bibinfo {year} {1984})}\BibitemShut {NoStop}%
\bibitem [{\citenamefont {Bos}\ \emph {et~al.}(2012)\citenamefont {Bos},
  \citenamefont {Chevillard}, \citenamefont {Scott},\ and\ \citenamefont
  {Rubinstein}}]{Bos2012}%
  \BibitemOpen
  \bibfield  {author} {\bibinfo {author} {\bibfnamefont {W.~J.~T.}\
  \bibnamefont {Bos}}, \bibinfo {author} {\bibfnamefont {L.}~\bibnamefont
  {Chevillard}}, \bibinfo {author} {\bibfnamefont {J.~F.}\ \bibnamefont
  {Scott}}, \ and\ \bibinfo {author} {\bibfnamefont {R.}~\bibnamefont
  {Rubinstein}},\ }\bibfield  {title} {\enquote {\bibinfo {title} {Reynolds
  number effect on the velocity increment skewness in isotropic turbulence},}\
  }\href@noop {} {\bibfield  {journal} {\bibinfo  {journal} {Physics of
  Fluids}\ }\textbf {\bibinfo {volume} {24}},\ \bibinfo {pages} {015108}
  (\bibinfo {year} {2012})}\BibitemShut {NoStop}%
\bibitem [{\citenamefont {Mininni}\ and\ \citenamefont
  {Pouquet}(2009)}]{Mininni2009}%
  \BibitemOpen
  \bibfield  {author} {\bibinfo {author} {\bibfnamefont {P.D.}\ \bibnamefont
  {Mininni}}\ and\ \bibinfo {author} {\bibfnamefont {A.}~\bibnamefont
  {Pouquet}},\ }\bibfield  {title} {\enquote {\bibinfo {title} {Helicity
  cascades in rotating turbulence},}\ }\href@noop {} {\bibfield  {journal}
  {\bibinfo  {journal} {Physical Review E}\ }\textbf {\bibinfo {volume} {79}},\
  \bibinfo {pages} {026304} (\bibinfo {year} {2009})}\BibitemShut {NoStop}%
\bibitem [{\citenamefont {Celani}\ \emph {et~al.}(2010)\citenamefont {Celani},
  \citenamefont {Musacchio},\ and\ \citenamefont {Vincenzi}}]{Celani2010}%
  \BibitemOpen
  \bibfield  {author} {\bibinfo {author} {\bibfnamefont {A.}~\bibnamefont
  {Celani}}, \bibinfo {author} {\bibfnamefont {S.}~\bibnamefont {Musacchio}}, \
  and\ \bibinfo {author} {\bibfnamefont {D.}~\bibnamefont {Vincenzi}},\
  }\bibfield  {title} {\enquote {\bibinfo {title} {Turbulence in more than two
  and less than three dimensions},}\ }\href@noop {} {\bibfield  {journal}
  {\bibinfo  {journal} {Physical Review Letters}\ }\textbf {\bibinfo {volume}
  {104}},\ \bibinfo {pages} {184506} (\bibinfo {year} {2010})}\BibitemShut
  {NoStop}%
\bibitem [{\citenamefont {Xia}\ \emph {et~al.}(2011)\citenamefont {Xia},
  \citenamefont {Byrne}, \citenamefont {Falkovich},\ and\ \citenamefont
  {Shats}}]{Xia2011}%
  \BibitemOpen
  \bibfield  {author} {\bibinfo {author} {\bibfnamefont {H.}~\bibnamefont
  {Xia}}, \bibinfo {author} {\bibfnamefont {D.}~\bibnamefont {Byrne}}, \bibinfo
  {author} {\bibfnamefont {G.}~\bibnamefont {Falkovich}}, \ and\ \bibinfo
  {author} {\bibfnamefont {M.}~\bibnamefont {Shats}},\ }\bibfield  {title}
  {\enquote {\bibinfo {title} {Upscale energy transfer in thick turbulent fluid
  layers},}\ }\href@noop {} {\bibfield  {journal} {\bibinfo  {journal} {Nature
  Physics}\ }\textbf {\bibinfo {volume} {7}},\ \bibinfo {pages} {321} (\bibinfo
  {year} {2011})}\BibitemShut {NoStop}%
\bibitem [{\citenamefont {Godeferd}\ and\ \citenamefont
  {Moisy}(2015)}]{Godeferd2015}%
  \BibitemOpen
  \bibfield  {author} {\bibinfo {author} {\bibfnamefont {F.~S.}\ \bibnamefont
  {Godeferd}}\ and\ \bibinfo {author} {\bibfnamefont {F.}~\bibnamefont
  {Moisy}},\ }\bibfield  {title} {\enquote {\bibinfo {title} {Structure and
  dynamics of rotating turbulence: A review of recent experimental and
  numerical results},}\ }\href@noop {} {\bibfield  {journal} {\bibinfo
  {journal} {Appl. Mech. Rev.}\ }\textbf {\bibinfo {volume} {67}},\ \bibinfo
  {pages} {030802} (\bibinfo {year} {2015})}\BibitemShut {NoStop}%
\bibitem [{\citenamefont {Biferale}\ \emph {et~al.}(2016)\citenamefont
  {Biferale}, \citenamefont {Bonaccorso}, \citenamefont {Mazzitelli},
  \citenamefont {van Hinsberg}, \citenamefont {Lanotte}, \citenamefont
  {Musacchio}, \citenamefont {Perlekar},\ and\ \citenamefont
  {Toschi}}]{Biferale2016}%
  \BibitemOpen
  \bibfield  {author} {\bibinfo {author} {\bibfnamefont {L.}~\bibnamefont
  {Biferale}}, \bibinfo {author} {\bibfnamefont {F.}~\bibnamefont
  {Bonaccorso}}, \bibinfo {author} {\bibfnamefont {I.~M.}\ \bibnamefont
  {Mazzitelli}}, \bibinfo {author} {\bibfnamefont {M.~A.~T.}\ \bibnamefont {van
  Hinsberg}}, \bibinfo {author} {\bibfnamefont {A.~S.}\ \bibnamefont
  {Lanotte}}, \bibinfo {author} {\bibfnamefont {S.}~\bibnamefont {Musacchio}},
  \bibinfo {author} {\bibfnamefont {P.}~\bibnamefont {Perlekar}}, \ and\
  \bibinfo {author} {\bibfnamefont {F.}~\bibnamefont {Toschi}},\ }\bibfield
  {title} {\enquote {\bibinfo {title} {Coherent structures and extreme events
  in rotating multiphase turbulent flows},}\ }\href@noop {} {\bibfield
  {journal} {\bibinfo  {journal} {Phys. Rev. X}\ }\textbf {\bibinfo {volume}
  {6}},\ \bibinfo {pages} {041036} (\bibinfo {year} {2016})}\BibitemShut
  {NoStop}%
\bibitem [{\citenamefont {Chertkov}\ \emph {et~al.}(2007)\citenamefont
  {Chertkov}, \citenamefont {Connaughton}, \citenamefont {Kolokolov},\ and\
  \citenamefont {Lebedev}}]{Chertkov2007}%
  \BibitemOpen
  \bibfield  {author} {\bibinfo {author} {\bibfnamefont {M.}~\bibnamefont
  {Chertkov}}, \bibinfo {author} {\bibfnamefont {C.}~\bibnamefont
  {Connaughton}}, \bibinfo {author} {\bibfnamefont {I.}~\bibnamefont
  {Kolokolov}}, \ and\ \bibinfo {author} {\bibfnamefont {V.}~\bibnamefont
  {Lebedev}},\ }\bibfield  {title} {\enquote {\bibinfo {title} {Dynamics of
  energy condensation in two-dimensional turbulence},}\ }\href@noop {}
  {\bibfield  {journal} {\bibinfo  {journal} {Physical Review Letters}\
  }\textbf {\bibinfo {volume} {99}},\ \bibinfo {pages} {084501} (\bibinfo
  {year} {2007})}\BibitemShut {NoStop}%
\bibitem [{\citenamefont {Dombre}\ \emph {et~al.}(1986)\citenamefont {Dombre},
  \citenamefont {Frisch}, \citenamefont {Greene}, \citenamefont {H\'enon},
  \citenamefont {Mehr},\ and\ \citenamefont {Soward}}]{Dombre1986}%
  \BibitemOpen
  \bibfield  {author} {\bibinfo {author} {\bibfnamefont {T.}~\bibnamefont
  {Dombre}}, \bibinfo {author} {\bibfnamefont {U.}~\bibnamefont {Frisch}},
  \bibinfo {author} {\bibfnamefont {J.~M.}\ \bibnamefont {Greene}}, \bibinfo
  {author} {\bibfnamefont {M.}~\bibnamefont {H\'enon}}, \bibinfo {author}
  {\bibfnamefont {A.}~\bibnamefont {Mehr}}, \ and\ \bibinfo {author}
  {\bibfnamefont {A.~M.}\ \bibnamefont {Soward}},\ }\bibfield  {title}
  {\enquote {\bibinfo {title} {Chaotic streamlines in the abc flows},}\
  }\href@noop {} {\bibfield  {journal} {\bibinfo  {journal} {Journal of Fluid
  Mechanics}\ }\textbf {\bibinfo {volume} {167}},\ \bibinfo {pages} {353–391}
  (\bibinfo {year} {1986})}\BibitemShut {NoStop}%
\bibitem [{\citenamefont {Moffatt}(2014)}]{Moffatt2014}%
  \BibitemOpen
  \bibfield  {author} {\bibinfo {author} {\bibfnamefont {H.~K.}\ \bibnamefont
  {Moffatt}},\ }\bibfield  {title} {\enquote {\bibinfo {title} {Helicity and
  singular structures in fluid dynamics},}\ }\href@noop {} {\bibfield
  {journal} {\bibinfo  {journal} {Proceedings of the National Academy of
  Sciences}\ }\textbf {\bibinfo {volume} {111}},\ \bibinfo {pages} {3663--3670}
  (\bibinfo {year} {2014})}\BibitemShut {NoStop}%
\bibitem [{\citenamefont {Sahoo}\ and\ \citenamefont
  {Biferale}(2015)}]{Sahoo2015bis}%
  \BibitemOpen
  \bibfield  {author} {\bibinfo {author} {\bibfnamefont {G.}~\bibnamefont
  {Sahoo}}\ and\ \bibinfo {author} {\bibfnamefont {L.}~\bibnamefont
  {Biferale}},\ }\bibfield  {title} {\enquote {\bibinfo {title} {Disentangling
  the triadic interactions in navier-stokes equations},}\ }\href@noop {}
  {\bibfield  {journal} {\bibinfo  {journal} {The European Physical Journal E}\
  }\textbf {\bibinfo {volume} {38}},\ \bibinfo {pages} {114} (\bibinfo {year}
  {2015})}\BibitemShut {NoStop}%
\bibitem [{\citenamefont {Linkmann}\ \emph {et~al.}(2017)\citenamefont
  {Linkmann}, \citenamefont {Sahoo}, \citenamefont {McKay}, \citenamefont
  {Berera},\ and\ \citenamefont {Biferale}}]{Linkmann2017}%
  \BibitemOpen
  \bibfield  {author} {\bibinfo {author} {\bibfnamefont {M.}~\bibnamefont
  {Linkmann}}, \bibinfo {author} {\bibfnamefont {G.}~\bibnamefont {Sahoo}},
  \bibinfo {author} {\bibfnamefont {M.}~\bibnamefont {McKay}}, \bibinfo
  {author} {\bibfnamefont {A.}~\bibnamefont {Berera}}, \ and\ \bibinfo {author}
  {\bibfnamefont {L.}~\bibnamefont {Biferale}},\ }\bibfield  {title} {\enquote
  {\bibinfo {title} {Effects of magnetic and kinetic helicities on the growth
  of magnetic fields in laminar and turbulent flows by helical fourier
  decomposition},}\ }\href@noop {} {\bibfield  {journal} {\bibinfo  {journal}
  {The Astrophysical Journal}\ }\textbf {\bibinfo {volume} {836}},\ \bibinfo
  {pages} {26} (\bibinfo {year} {2017})}\BibitemShut {NoStop}%
\end{thebibliography}%

\end{document}